\pdfoutput=1
\documentclass{article}

\PassOptionsToPackage{numbers, compress}{natbib}

\usepackage[main, final]{neurips_2026}

\usepackage[export]{adjustbox}
\usepackage[ruled]{algorithm2e}
\usepackage[inline, shortlabels]{enumitem}
\usepackage[T1]{fontenc}
\usepackage{hyperref}
\usepackage{microtype}
\usepackage{pifont}
\usepackage[table]{xcolor}
\usepackage{xurl}
\usepackage{graphicx}
\usepackage{booktabs}
\usepackage{tabularray}
\usepackage{listings}
\usepackage{amsmath, amsfonts}
\usepackage{nicefrac}

\usepackage{tabularx}
\usepackage{graphicx}

\usepackage{wrapfig}

\UseTblrLibrary{booktabs}

\newcommand{\R}{\mathbb{R}}

\newcommand{\e}{\begin{equation}}
\newcommand{\ee}{\end{equation}}
\newcommand{\en}{\begin{equation*}}
\newcommand{\een}{\end{equation*}}
\newcommand{\eqn}{\begin{eqnarray}}
\newcommand{\eeqn}{\end{eqnarray}}
\newcommand{\bmat}{\begin{bmatrix}}
\newcommand{\emat}{\end{bmatrix}}

\DeclareMathAlphabet\mathbfcal{OMS}{cmsy}{b}{n}

\newcommand{\mrm}{\mathrm}

\newcommand{\vct}[1]{\boldsymbol{#1}}
\newcommand{\mtx}[1]{\boldsymbol{#1}}

\newcommand{\T}{\mathrm{T}}

\DeclareMathOperator*{\argmax}{\text{arg~max}}

\newcommand{\calD}{\mathcal{D}}

\newcommand{\calH}{\mathcal{H}}

\newcommand{\calL}{\mathcal{L}}

\newcommand{\calO}{\mathcal{O}}

\newcommand{\calT}{\mathcal{T}}

\newcommand{\vg}{\vct{g}}

\newcommand{\vl}{\vct{l}}

\newcommand{\vx}{\vct{x}}
\newcommand{\vy}{\vct{y}}
\newcommand{\vz}{\vct{z}}
\newcommand{\valpha}{\vct{\alpha}}

\newcommand{\mB}{\mtx{B}}

\newcommand{\mF}{\mtx{F}}
\newcommand{\mG}{\mtx{G}}

\newcommand{\mK}{\mtx{K}}

\newcommand{\mP}{\mtx{P}}
\newcommand{\mQ}{\mtx{Q}}

\newcommand{\mV}{\mtx{V}}
\newcommand{\mW}{\mtx{W}}
\newcommand{\mX}{\mtx{X}}

\newcommand{\mZ}{\mtx{Z}}

\makeatletter
\newcommand{\ssymbol}[1]{\@fnsymbol{#1}}
\newcommand{\romanNumeral}[1]{\expandafter\@slowromancap\romannumeral #1@}
\makeatother

\definecolor{Gray}{gray}{0.9}
\definecolor{lgray}{gray}{0.95}
\definecolor{LightCyan}{rgb}{0.92,0.92,1}
\definecolor{lyellow}{rgb}{1,1,0.92}
\definecolor{lgreen}{rgb}{0.92,1,0.95}
\definecolor{llblue}{rgb}{0.92,0.93,0.95}
\definecolor{lred}{rgb}{1,0.85, 0.85}
\definecolor{lblue}{rgb}{0.92,0.95,1}
\definecolor{tabhighlight}{HTML}{e5e5e5}

\usepackage[utf8]{inputenc} %
\usepackage[T1]{fontenc}    %
\usepackage{hyperref}       %
\usepackage{url}            %
\usepackage{booktabs}       %
\usepackage{amsfonts}       %
\usepackage{nicefrac}       %
\usepackage{microtype}      %
\usepackage{xcolor}         %

\usepackage{amssymb}
\usepackage{multirow}
\usepackage{subcaption}
\usepackage{siunitx}
\usepackage{makecell}
\usepackage{xspace}

\newcommand{\method}{CART\xspace}
\newcommand{\eg}{e.g.,\xspace}

\newcommand{\td}[1]{\textcolor{black}{#1}}

\usepackage{cleveref}

\title{Learning Where and What to Restore for Composite Image Restoration}

\author{%
  Jiachen Jiang$^{1,*}$ \quad
  Tianyu Ding$^{2,*,\dagger}$ \quad
  Ke Zhang$^{3}$ \quad
  Jinxin Zhou$^{2}$ \\
  \textbf{Tianyi Chen$^{2}$} \quad
  \textbf{Ilya Zharkov$^{2}$} \quad
  \textbf{Zhihui Zhu$^{1}$} \quad
  \textbf{Luming Liang$^{2,\dagger}$} \\[4pt]
  $^{1}$The Ohio State University \quad
  $^{2}$Microsoft \quad
  $^{3}$Johns Hopkins University
}

\begin{document}
\maketitle

\begingroup
\renewcommand{\thefootnote}{\fnsymbol{footnote}}
\footnotetext[1]{Equal contribution. This work was done when Jiachen Jiang was an intern at Applied Sciences Group, Microsoft.}
\footnotetext[2]{Corresponding authors.}
\endgroup

\begin{figure}[h]
  \centering
  \includegraphics[width=\linewidth]{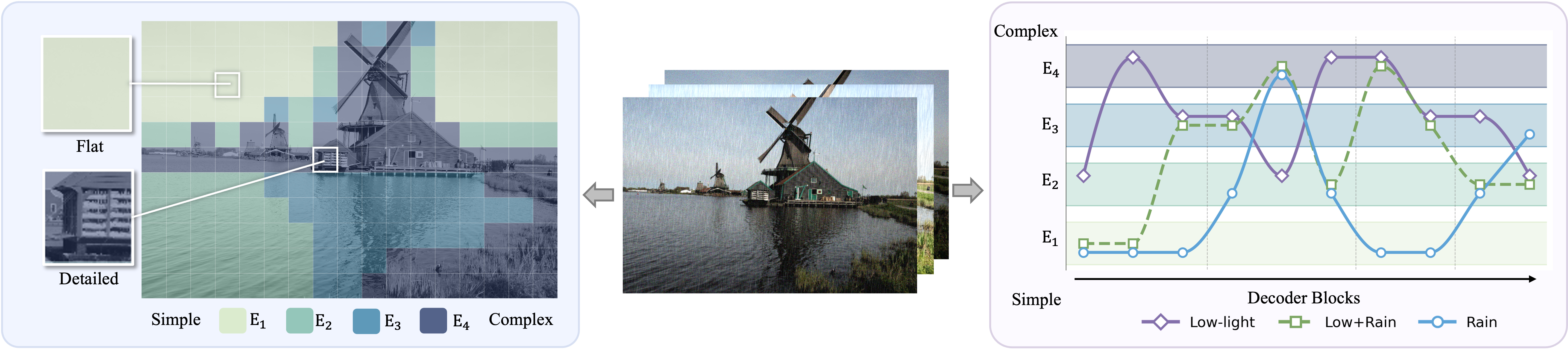}
  \caption{%
    \td{Composite image restoration requires jointly deciding
    \emph{where} and \emph{what} to restore.
    The degraded input (center) is processed along two complementary axes.
    \textbf{Left (where):} each spatial patch is routed to one of the capacity-graded experts (e.g. $E_1$ to $E_4$, simple to complex).
    Flat regions are handled by lightweight experts,
    while detailed regions activate higher-capacity ones.
    \textbf{Right~(what):} the expert chosen at each decoder block traces a curve
    across capacity levels; each elemental degradation (Low-light, Rain) follows
    its own characteristic trajectory, and the composite Low+Rain path (dashed)
    blends those of its single-degradation constituents.}}
  \label{fig:teaser}
\end{figure}

\begin{abstract}
\td{Real-world degraded images often contain multiple co-occurring degradation types, making composite image restoration fundamentally different from the single-degradation setting assumed by most existing all-in-one methods.
These methods typically apply uniform spatial computation and single-label task conditioning, limiting both spatial adaptivity and explicit modeling of degradation mixtures.
We propose \method (Composite-Adaptive Routing and Task Conditioning), a unified framework that jointly decides \emph{where} to spend computation and \emph{what} degradation cues should guide restoration: a spatial mixer routes patches to graded-capacity experts by local restoration difficulty, and a channel mixer routes each image through degradation-specific experts, conditioned on the global task feature with multi-hot classification supervision.
Empirically, \method achieves state-of-the-art performance for composite degradation restoration on CDD-11 and delivers the best results to date on the conventional 3-task and 5-task all-in-one benchmarks.}
\end{abstract}

\section{Introduction}
\label{sec:intro}
\begin{wrapfigure}{r}{0.5\linewidth}
    \vspace{-5pt}
    \centering
    \includegraphics[width=\linewidth]{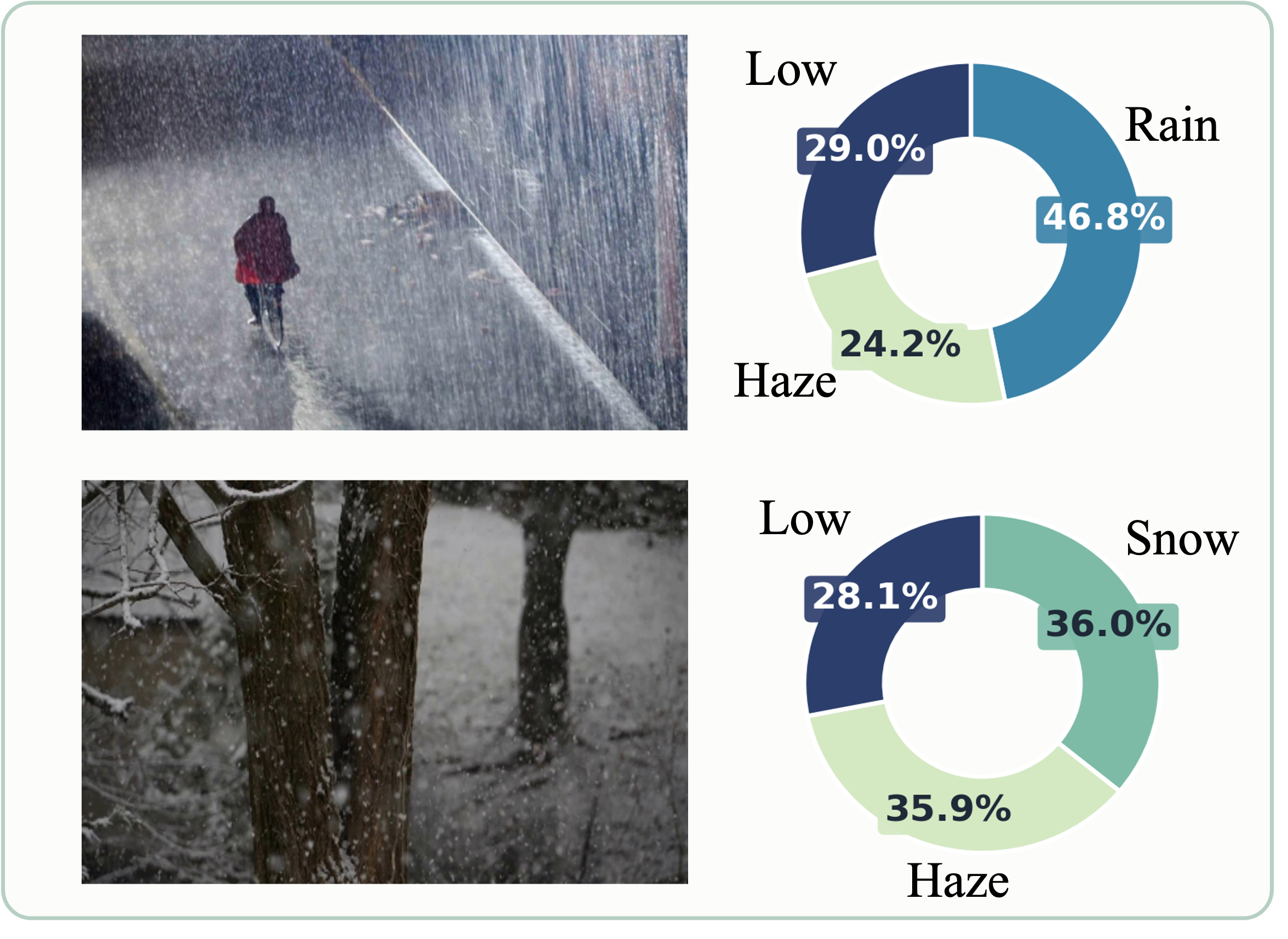}
    \caption{\td{Real-world samples~\citep{zhu2023learning} contain multiple co-occurring degradations rather than one dominant
  corruption; proportions are predicted by our channel mixer (\Cref{sec:task}).}}
    \label{fig:realworld_cart_mixer}
    \vspace{-5pt}
\end{wrapfigure}

\td{Real-world degraded images rarely contain a single isolated corruption.
Image restoration aims to recover clean images from degraded observations, and
recent restoration models have achieved strong results across diverse
settings \cite{liang2021swinir,Wang_2022_CVPR,zamir2022restormer,chen2022simple,zhou2024dream}. Yet most existing methods are developed for specific degradation types or
assume one dominant corruption, which limits their practical use in autonomous
driving, adverse-weather perception \cite{chen2023always,valanarasu2022transweather}
and night-time surveillance \cite{lamba2021restoring}, where low-light,
haze, rain, blur, and other degradations may co-exist
\cite{anwar2021attention,zamir2020cycleisp,zamir2020learning,guo2024onerestore,wang2025dpl}.
Figure~\ref{fig:realworld_cart_mixer} illustrates this phenomenon on a real-world dataset~\cite{zhu2023learning}, where rain, snow, haze, and low-light co-occur in varying proportions rather than one dominant corruption.
We refer to this setting as \emph{composite image restoration}. Compared with
the single-degradation setting assumed by most prior work, composite
restoration requires the model to disentangle overlapping artifacts with
different spatial scales and different restoration requirements.}

\vspace{-10pt}
\td{Recent all-in-one image restoration methods attempt to address multiple tasks
within a unified model
\cite{li2022airnet,potlapalli2023promptir,zhang2023ingredient,conde2024instructir,zamfir2025complexity,tian2025dfpir,luo2025visual}.
However, a key mismatch remains between their design assumptions and the
composite setting. Different degradations place different demands on the
network: denoising and deraining emphasize precise local detail processing,
whereas deblurring, dehazing, and low-light enhancement depend more strongly
on broad spatial context and image-level statistics \cite{xu2024unified}.
Still, representative architectures largely apply a shared computation pattern
across spatial regions rather than allocating computation according to local
restoration difficulty
\cite{li2022airnet,potlapalli2023promptir,conde2024instructir,tian2025dfpir,luo2025visual}.
Moreover, many conditioning-based methods summarize the degradation with one global
signal, such as an image-level prompt or instruction
\cite{potlapalli2023promptir,wang2023promptrestorer,conde2024instructir,ai2024multimodal,luo2025visual},
which is insufficient when multiple degradation types are present at once. Together, these limitations indicate that current composite restoration methods lack joint spatial and task adaptivity.}

\td{We address these limitations with \method, a unified restoration framework that
adapts along two complementary axes: \emph{where} the model spends computation, and
\emph{what} degradation types it specializes for.
A \emph{spatial mixer} routes patches to experts of graded capacity by local restoration
 difficulty.
A \emph{channel mixer} predicts a multi-hot degradation vector at the bottleneck and uses
 the pre-classifier feature to route each image through degradation-specialized channel
 experts at every decoder block.
The resulting model adapts both spatially and across degradation types to mixed
degradations without requiring manual task specification at inference time.}

\td{We evaluate \method on CDD-11 \cite{guo2024onerestore}, a benchmark that covers single,
double, and triple combinations of low-light, haze, rain, and snow. We
further evaluate it on the conventional 3-task and 5-task all-in-one benchmarks spanning denoising,
deraining, dehazing, deblurring, and low-light enhancement. Across these settings, \method achieves state-of-the-art quantitative
results and clearer visual reconstructions, while remaining the most efficient among compared methods in FLOPs and runtime.}

\td{We summarize our contributions as follows:
\begin{itemize} [leftmargin=1.5em,itemsep=2pt,topsep=2pt]
  \item We formulate composite image restoration as requiring two complementary forms of adaptivity: spatial adaptivity to allocate computation by local difficulty, and task adaptivity to disentangle the compositional identity of co-occurring degradations.
  \item We propose \method, a unified framework that combines patch-wise content-adaptive routing with task-conditioned channel routing driven by a multi-hot degradation prediction. The former allocates computation by local restoration difficulty, while the latter routes each image through degradation-specialized channel experts within a single network.
  \item We validate \method on CDD-11 and the conventional 3-task and 5-task all-in-one benchmarks. \method achieves state-of-the-art results on both composite and standard multi-degradation settings while remaining the most efficient among compared methods in FLOPs and runtime.
\end{itemize}}

\section{Related work}
\label{sec:related}

\td{Prior work on all-in-one restoration is most related to our setting along two axes: how the model represents degradation identity, and how it allocates computation. Composite restoration requires both, since the model must infer which degradations co-occur and decide where to intensify processing.}

\td{\textbf{All-in-one restoration and task conditioning.}
All-in-one restoration methods seek a single model that can cover multiple degradation types.
Early approaches such as IPT~\cite{chen2021IPT} and AirNet~\cite{li2022airnet} rely on a shared backbone to handle several tasks, while later methods inject task cues through prompts, instructions, or auxiliary descriptors, as in PromptIR~\cite{potlapalli2023promptir}, PromptRestorer~\cite{wang2023promptrestorer}, IDR~\cite{zhang2023ingredient}, InstructIR~\cite{conde2024instructir}, multimodal conditioning~\cite{ai2024multimodal}, DeFusion~\cite{luo2025visual} and AdaIR~\cite{cui2025adair}.
These methods broaden the scope of all-in-one restoration, but most summarize the input with a single global condition or treat degradation cues independently.
Such designs are effective when one degradation dominates, yet they are less well matched to composite inputs, where multiple degradation types may be active simultaneously and must be represented jointly rather than collapsed into one image-level prompt.}

\td{\textbf{Adaptive computation.}
Dynamic computation~\cite{bengio2013estimating,shazeer2017outrageously} allocates variable capacity according to input complexity.
In restoration, MGA~\cite{hu2022restore} uses a mask predictor to select under-resolved patches for a dedicated refine-net, Path-Restore~\cite{yu2021path} routes local regions through different depths, CAMixerSR~\cite{wang2024camixersr} classifies patches as hard or easy for super-resolution, and MoCE-IR~\cite{zamfir2025complexity} routes whole images to experts of different complexity and receptive field.
These methods confirm that restoration benefits from content-dependent computation, but their routing granularity is typically binary (skip/refine or easy/hard), and they address only \emph{where} more computation should be spent without explicitly modeling \emph{what} degradation types are present.}

\td{\textbf{Composite degradation restoration.}
Composite restoration differs from generic all-in-one restoration in that the degradation label is itself \emph{a set} rather than a single category, requiring the model to represent and disentangle co-occurring artifacts.
Methods such as TransWeather~\cite{valanarasu2022transweather}, WeatherDiff~\cite{ozdenizci2023restoring}, and WGWSNet~\cite{zhu2023learning} address multiple adverse-weather conditions, but they focus on a limited weather family and typically treat each image as having one dominant type.
More directly related to our setting, OneRestore~\cite{guo2024onerestore} introduces the CDD-11 benchmark covering mixtures of low-light, haze, rain, and snow, and conditions restoration on scene-level descriptors; AllRestorer~\cite{mao2024allrestorer} similarly relies on externally derived scene attributes; DPL~\cite{wang2025dpl} explores dual-level prototypes for composite inputs.
These methods require additional descriptors to identify the active degradation mixture.
BioIR~\cite{cuibio} avoids external descriptors by using a bio-inspired architecture, but still lacks explicit task identification.
DCPT~\cite{hu2025universal} takes a different route by using degradation classification as a pre-training objective, yet the classification signal does not directly control the restoration pathway.
\method instead couples both requirements: a multi-hot classification head infers the active degradation set and feeds it into task-conditioned channel routing, while patch-level spatial routing with a graded capacity spectrum allocates computation according to local restoration difficulty.}

\section{Method}
\label{sec:method}

\td{Let $\calT=\{T_i\}_{i=1}^M$ denote a set of $M$ degradation types,
and let $\calD \subseteq \calT$ ($|\calD| \geq 1$) be the unknown active subset.
Given a clean image $I$, a compositely degraded observation $I_\calD$ is modeled as
\e
  I_\calD(x) \;=\; \calH_{\calD}\!\bigl(I;\, x\bigr),
  \label{eq:degradation}
\ee
where $x$ indexes spatial location and $\calH_\calD$ denotes the joint degradation operator
induced by $\calD$, whose effect varies across the image.
The severity of $\calH_\calD$ is \emph{spatially heterogeneous}: recovering fine textures demands more capacity than correcting flat regions, while the active subset $\calD$ is itself \emph{unobserved} at inference. We seek a single model $f_\theta$ such that $\hat{I} = f_\theta(I_\calD) \approx I$
without access to either a severity map or the degradation label $\calD$.}

\begin{figure}[t]
\centering
\includegraphics[width=0.99\linewidth]{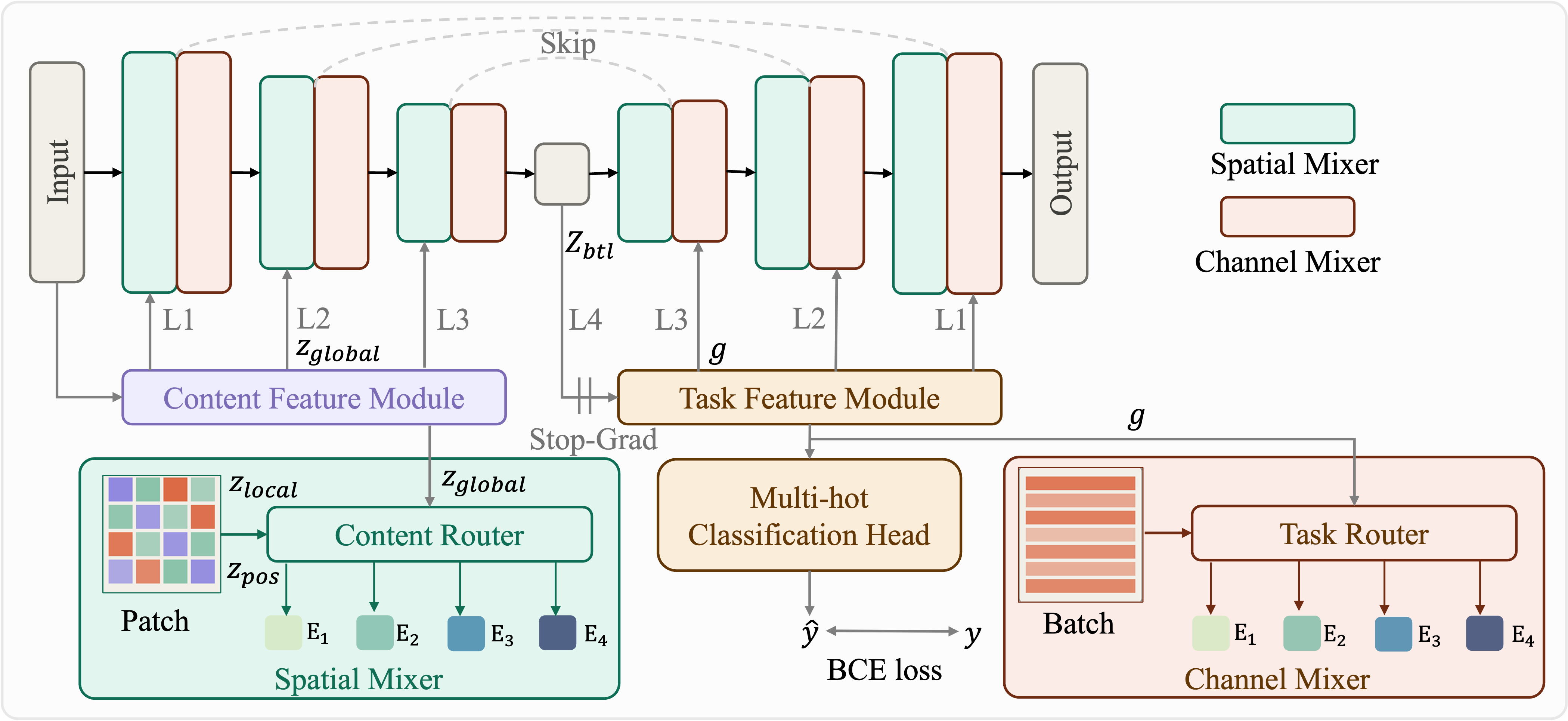}
\caption{\td{Overview of \method.
  A U-Net encoder-decoder alternates spatial and channel mixer blocks.
  The \emph{spatial mixer} routes each patch to one of $K$ experts of graded capacity by local restoration difficulty.
  A task feature module extracts a global task feature $\vg$ at the bottleneck and supervises it with multi-hot degradation classification;
  the \emph{channel mixer} then uses this feature to route each image
  through one of $K$ channel-attention experts in the decoder.}}
\label{fig:overview}
\end{figure}

\td{Two challenges follow directly from~\eqref{eq:degradation}. First, the effective severity of $\calH_\calD$ varies across spatial locations~$x$:
a uniform sky requires minimal computation to restore, while a building with dense edges and fine detail demands substantially more. Applying a fixed computational budget uniformly across all regions
inevitably wastes capacity on simple areas and starves complex ones.
Second, the active subset $\calD$ is unknown at inference
and combinatorial in nature---$M$ elemental types yield $2^M - 1$
possible degradation mixtures, each requiring qualitatively different processing.
Without an explicit mechanism to disentangle which degradations are present,
a shared set of parameters is forced to compromise across tasks
rather than specialize.}

\td{CART introduces two complementary mixer modules to tackle these challenges as shown in \Cref{fig:overview}. A \emph{spatial mixer} splits features into patches and routes each patch to one of $K$ spatial attention modules of increasing capacity, spending more computation on complex regions and less on simple ones (\Cref{sec:content}). A \emph{channel mixer} uses a predicted multi-hot vector $\hat{\vy} \in [0,1]^M$ to route each image through different channel attention modules, so that different degradation types activate different parameters instead of competing over shared ones (\Cref{sec:task}). The two mixers alternate within the backbone, letting the model jointly refine \emph{where} computation is allocated and
 \emph{what} degradation cues guide restoration at each feature scale.}

\subsection{Spatial mixer: where to allocate computation}
\label{sec:content}

\td{In composite-degraded images, the spatial distribution of restoration difficulty is more heterogeneous than in the single-degradation case: rain-streaked edges, haze-covered flat regions, and areas where both overlap present qualitatively different demands within the same image.
Prior spatially adaptive methods typically make binary routing decisions (skip/refine or easy/hard), which cannot capture this finer-grained variation in difficulty.
We propose a spatial mixer in which each block contains $K$ experts forming a graded capacity spectrum, and a lightweight router assigns each patch to exactly one expert based on local restoration difficulty.}

\textbf{Complexity experts.}
\td{The $K$ experts are windowed self-attention blocks that share the same attention structure but differ in hidden dimension~$d_k$.
We space $\{d_k\}_{k=0}^{K-1}$ linearly from $d/K$ to~$d$, where $d$ is the block (residual) channel dimension, forming $K$ ordered capacity levels rather than independent or uniform experts.
This graded design lets the model match each patch to the minimal sufficient expert: simple regions (\eg uniform sky) are handled by a low-dimensional expert, while complex regions (\eg textured edges under mixed degradations) receive full-capacity attention.
Selecting the appropriate expert for each patch, however, requires information beyond its local feature alone, which motivates the router design below.}

\textbf{Content router.}
\td{The router must judge patch-level restoration difficulty, yet features in deeper encoder layers have progressively lost the local spatial detail.
We address this by supplementing each patch's local representation with a content feature module drawn from first-level encoder features, which still retain fine-grained structural information.
Concretely, the router receives three inputs:
(i)~a \emph{local embedding} $\vz_{\mrm{local}} \in \R^{C_\ell}$,
    obtained by average-pooling the residual stream features within each  window;
(ii)~a \emph{global content descriptor} $\vz_{\mrm{global}} \in \R^{C_g}$,
    extracted from first-level encoder features
    and interpolated to each window position; and
(iii)~a \emph{positional encoding} $\vz_{\mrm{pos}} \in \R^2$,
    encoding normalized spatial coordinates.
A two-layer MLP maps their concatenation to $K$ routing logits:
\e
  \vl = \mathrm{MLP}([\vz_{\mrm{local}},\;
         \vz_{\mrm{global}},\;
         \vz_{\mrm{pos}}]) \;\in\; \R^K.
  \label{eq:router}
\ee
At inference, each patch is hard-assigned to $\argmax_k l_k$.
During training, Gumbel-Softmax~\cite{jang2016categorical} makes this discrete assignment differentiable, and a standard load-balancing loss~\cite{fedus2022switch} prevents the router from collapsing to a single expert (details in \Cref{sec:app_arch}).}

\subsection{Channel mixer: what degradation cues guide restoration}
\label{sec:task}

\td{In composite-degraded images, the active degradation set $\calD$ is unobserved at inference and varies compositionally: an image corrupted by haze{+}rain is related to both haze-only and rain-only inputs, yet distinct from either.
Prior all-in-one methods condition restoration on a single global cue such as a one-hot label, text prompt, or scene descriptor, which treats every mixture as an indivisible category and discards this compositional structure.
We propose a channel mixer that uses a global task feature $\vg$, extracted by
a task feature module at the bottleneck and supervised by multi-hot degradation
classification, to route each image through one of $K$ channel-attention experts.}

\textbf{Task-aware routing signal.} \td{A task feature module extracts a global task feature $\vg$ from the bottleneck feature $\mZ_{\mrm{btl}}$, but $\vg$ inherits the entanglement of content, degradation, and reconstruction cues from $\mZ_{\mrm{btl}}$ and provides no \emph{prior} for what the channel router needs.
We therefore attach a classification head on top of $\vg$ and supervise it with a degradation-recognition objective; under this classification supervision, $\vg$ is shaped toward the neural-collapse geometry~\cite{papyan2020prevalence, zhu2021geometric} in which same-class features converge and different classes separate, so that $\vg$ encodes task information.
We apply a stop-gradient on $\mZ_{\mrm{btl}}$ before the task feature module, so the classification objective does not back-propagate into the shared encoder and the bottleneck feature retains its reconstruction signal.}

\begin{wrapfigure}{r}{0.4\textwidth}
  \vspace{-15pt}
    \centering
    \includegraphics[width=.99\linewidth]{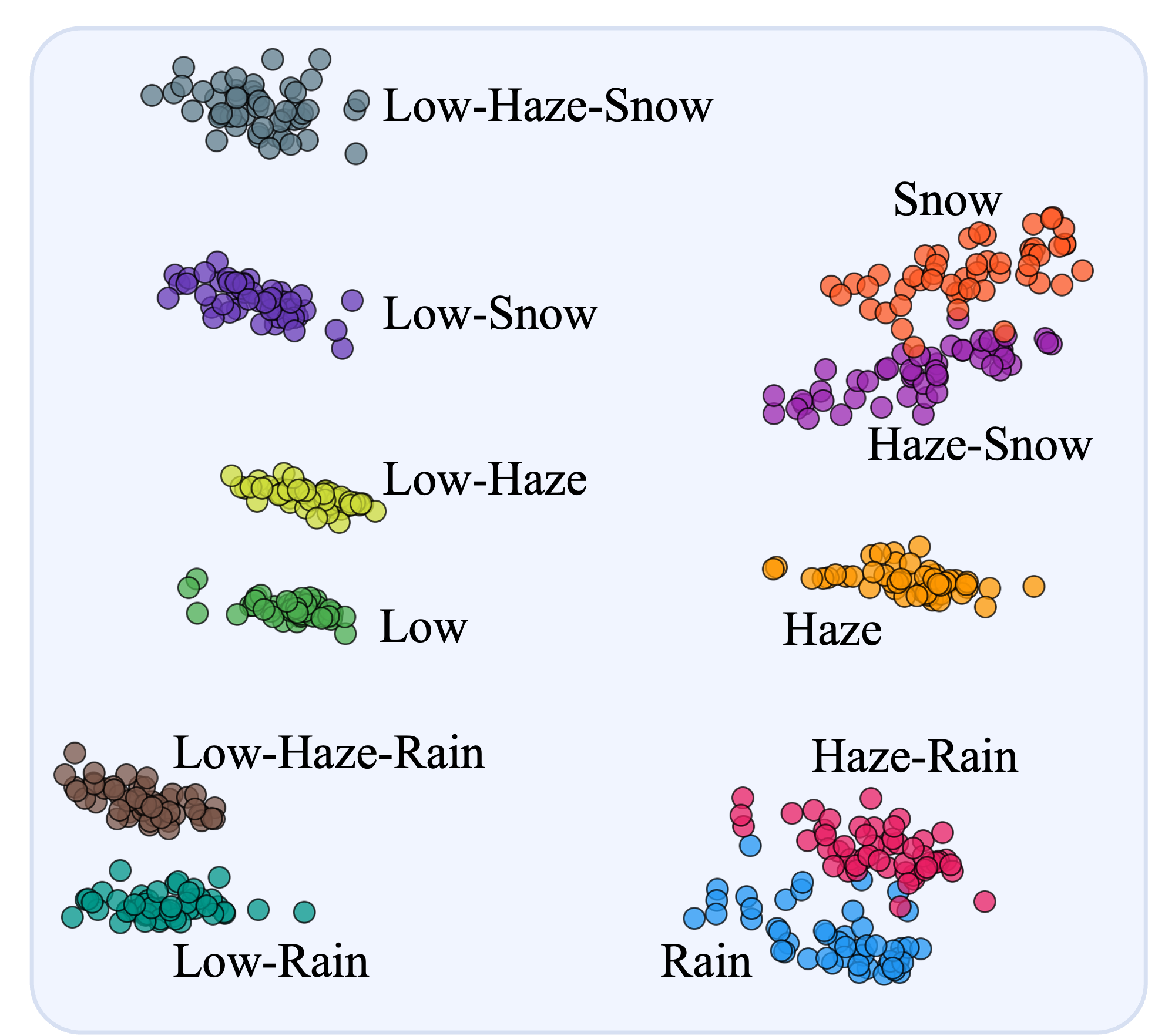}
 \caption{\td{PCA of the task feature $\vg$. Composite clusters lie between their single-degradation constituents.}
  }
  \label{fig:pca}
  \vspace{-5pt}
\end{wrapfigure}

\textbf{Compositional task representation.}
\td{A natural classifier would use one output per degradation combination, but this has two drawbacks for composite restoration: the number of categories grows as $2^M{-}1$, and each mixture is treated as an independent class, discarding the compositional structure that links, \eg haze{+}rain to both haze-only and rain-only.
We instead supervise $M$ independent outputs with binary cross-entropy, producing $\hat{\vy} \in [0,1]^M$.
Under this multi-hot supervision, the global task feature $\vg$ for a composite input aligns with components of its constituent single-degradation features, so parameter sharing across related tasks emerges as a property of the learned representation rather than a fixed architectural choice. A PCA projection of $\vg$ (\Cref{fig:pca}) supports this: across CDD-11 conditions, clusters for different
degradations are well separated, and each composite-degradation cluster lies between its constituent single-degradation
  ones.}

\textbf{Channel experts and image-level routing.}
\td{Unlike spatial difficulty, which varies within an image, the active degradation set $\calD$ is a global property of the image, so we route at the image level rather than per patch.
A task router takes the concatenation of $\vg$ and the global-average-pooled feature map and produces $K$ logits; hard $\argmax$ is used at inference, with Gumbel-Softmax~\cite{jang2016categorical} and a standard load-balancing loss during training (same machinery as the spatial mixer; details in \Cref{sec:app_arch}).
The $K$ experts are channel-attention blocks sharing the same graded capacity schedule as the spatial experts (hidden dimensions from $d/K$ to~$d$), but operating along the channel axis, so the spatial and channel mixers are complementary in the dimension they modulate.}

\td{Because this routing is applied at every decoder block, each elemental degradation induces a distinct routing path, and composite inputs produce paths that blend the constituent patterns.
This yields task-specific specialization within a shared network without separate branches or task labels at inference, so different degradations activate their own parameter subsets rather than competing for a shared pool (visualized in \Cref{sec:routing_vis}).}

\subsection{Training objective}
\label{sec:loss}

\td{The training objective combines reconstruction losses on the output image with auxiliary losses that supervise the two mixer mechanisms:
\e
\calL = \calL_{\mrm{recon}}
      + \lambda_{\mrm{ms}}\,\calL_{\mrm{ms}}
      + \lambda_{\mrm{cdrl}}\,\calL_{\mrm{cdrl}}
      + \lambda_{\mrm{lb}}\,\calL_{\mrm{lb}}
      + \lambda_{\mrm{task}}\,\calL_{\mrm{task}}.
\label{eq:total_loss}
\ee
Restoration quality is supervised by three terms: a smooth-$\ell_1$ pixel loss $\calL_{\mrm{recon}}$, a multi-scale structural similarity loss $\calL_{\mrm{ms}}$~\cite{wang2003multiscale} that encourages perceptually consistent structure, and the composite-degradation contrastive loss $\calL_{\mrm{cdrl}}$~\cite{guo2024onerestore}, which contrasts the output against all degradation types rather than only those present in the input.
Two auxiliary terms supervise the mixers: $\calL_{\mrm{lb}}$ is a standard load-balancing loss~\cite{fedus2022switch} averaged over all routed stages (both spatial and channel), and $\calL_{\mrm{task}}$ is the binary cross-entropy on the multi-hot degradation prediction at the bottleneck.}

\begin{figure*}[t]
  \centering
  \includegraphics[width=\linewidth]{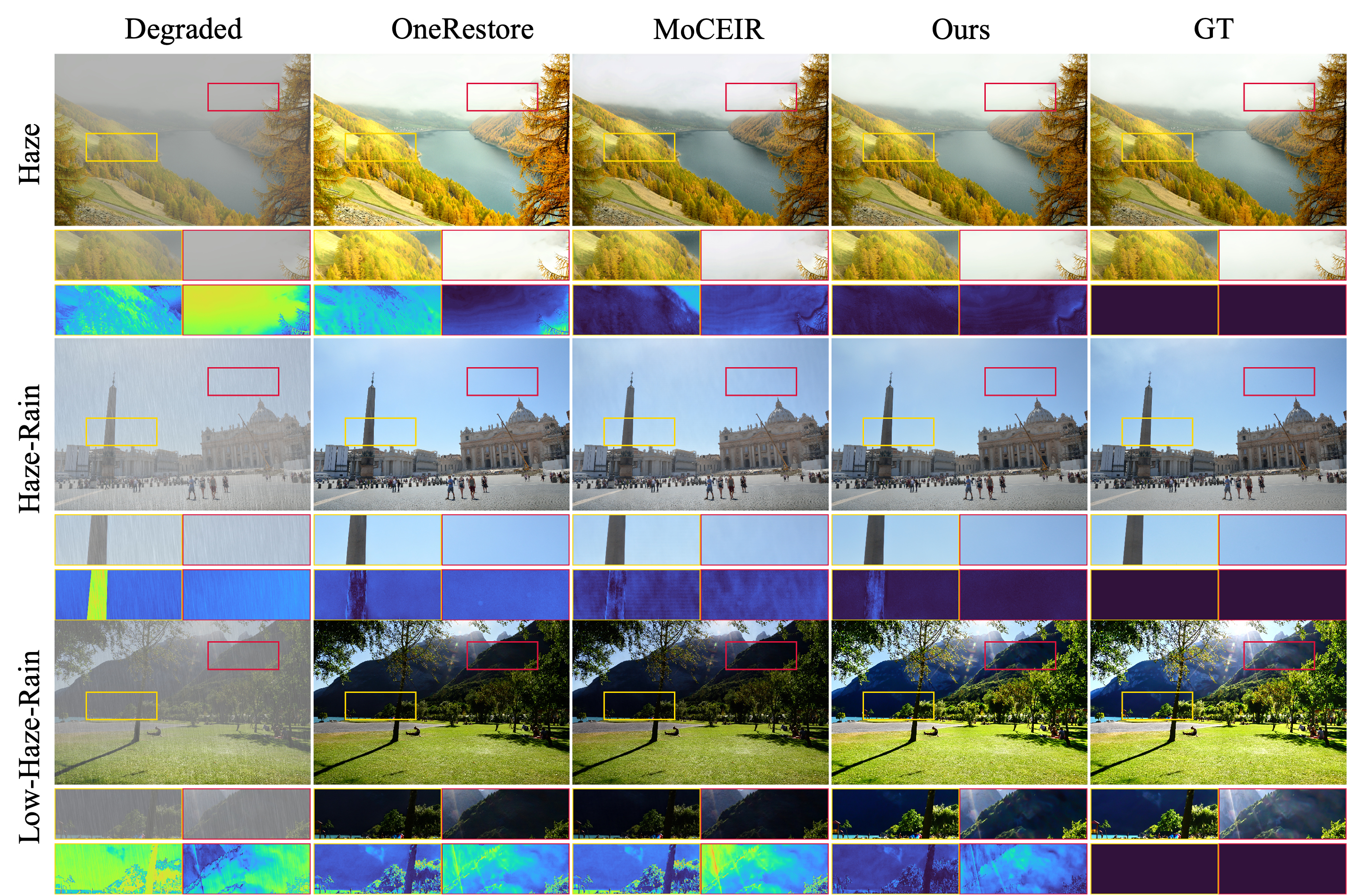}
  \caption{\td{Visual results on CDD-11. \method vs.\
  OneRestore~\cite{guo2024onerestore} and
  MoCE-IR~\cite{zamfir2025complexity} on three cases: haze, haze+rain,
  and low+haze+rain.
  For each case, the top strip shows the degraded input, the
  restorations from each method, and the ground truth;
  the middle strip zooms into regions;
  the bottom strip shows the per-pixel error map to ground truth
  (darker = smaller residual).}}
  \label{fig:visual-compare}
\end{figure*}

\section{Experiments}
\label{sec:experiments}

\subsection{Setup}
\label{sec:setup}

\td{\method targets \emph{composite} image restoration, where multiple
degradations co-occur in a single input.
We thus evaluate primarily on the composite setting, and additionally on
conventional all-in-one benchmarks to verify the design generalizes to
inputs with a single active degradation.}

\textbf{Datasets.}
\td{\emph{Composite restoration.}
CDD-11~\cite{guo2024onerestore} contains images with one to three
co-occurring degradations from low-light (L), haze (H), rain (R), and snow
(S), covering $11$ conditions in total: $4$ single, $5$ two-way, and $2$
three-way.
\emph{Conventional all-in-one (3D/5D).}
Following~\cite{potlapalli2023promptir}, the 3-task setting (3D) combines
Gaussian denoising on
BSD400~\cite{arbelaez2010contour}+WED~\cite{ma2016waterloo} with
$\sigma\in\{15,25,50\}$ (tested on BSD68~\cite{martin2001database}),
deraining on Rain100L~\cite{yang2017deep}, and dehazing on
SOTS~\cite{li2018benchmarking}; the 5-task setting (5D) adds deblurring on
GoPro~\cite{nah2017deep} and low-light enhancement on
LOL-v1~\cite{wei2018deep}.}

\textbf{Implementation details.}
\td{\method adopts a four-level U-Net with base channel width $d{=}48$ and
block counts $[2,3,3,4]$ from top to bottom.
Each routed layer uses $K{=}4$ experts whose hidden dimensions grow
linearly as $[12,24,36,48]$ at the base level and scale by $2^{\ell}$ at
level~$\ell$.
We train with Adam ($\beta_1{=}0.9$, $\beta_2{=}0.999$) at batch size $32$,
with an initial learning rate of $2{\times}10^{-4}$ and cosine annealing
after a $15$-epoch warmup.
Inputs are $128{\times}128$ random crops with horizontal and vertical
flips, and the Gumbel-Softmax temperature $\tau$ is annealed from $2.5$ to
$0.7$ over training.
Loss weights are
$\lambda_{\mrm{ms}}{=}0.3$,
$\lambda_{\mrm{cdrl}}{=}0.1$,
$\lambda_{\mrm{lb}}{=}0.01$, and
$\lambda_{\mrm{task}}{=}0.01$.
We train one unified model for $800$ epochs on CDD-11 and one for $120$
epochs on each of the 3D and 5D benchmarks.
We report PSNR~(dB) and SSIM, with a macro-average over the $11$ CDD-11
conditions and per-task averages on 3D/5D.}

\subsection{Main results}
\label{sec:main_results}

\textbf{Composite degradations.}
\td{On the composite setting, we evaluate \method on
CDD-11~\cite{guo2024onerestore} (\Cref{tab:cdd11}) against
representative all-in-one and composite restoration methods:
AirNet~\cite{li2022airnet}, PromptIR~\cite{potlapalli2023promptir},
WGWSNet~\cite{zhu2023learning},
WeatherDiff~\cite{ozdenizci2023restoring},
OneRestore~\cite{guo2024onerestore}, and the strongest prior baseline
MoCE-IR~\cite{zamfir2025complexity}.
\method attains the highest average PSNR of $29.95$~dB
($+0.90$~dB over MoCE-IR) and ranks best on $10$ of $11$ conditions
in PSNR and on all $11$ in SSIM.
The advantage widens on multi-degradation mixtures (\eg\
$+1.68$~dB on H+R and $+2.09$~dB on H+S over MoCE-IR), where
multi-hot task conditioning preserves the identities of co-occurring
degradations that single-cue baselines collapse.}

\textbf{Conventional all-in-one (3D and 5D).}
\td{To verify that the design also generalizes to inputs with a single
active degradation, we evaluate on the standard 3-task and 5-task
all-in-one benchmarks (\Cref{tab:allinone-3d,tab:allinone-5d}).
\method attains the best average PSNR on both ($+0.37$~dB on 3D and
$+0.29$~dB on 5D over MoCE-IR), with the largest gains on tasks
demanding fine local detail or large global statistics shifts:
heavy-noise denoising ($\sigma{=}50$: $+0.90$~dB on 3D) and
low-light enhancement ($+0.87$~dB on 5D).
On well-saturated tasks such as dehazing and deblurring,
\method is on par with the best prior method while remaining ahead on
average, consistent with our two-mixer design that adapts capacity
to local difficulty and specializes features per degradation type.}

\textbf{Visual results.}
\td{\Cref{fig:visual-compare} compares \method to OneRestore and
MoCE-IR on three CDD-11 cases of increasing difficulty.
On haze, OneRestore introduces a color cast and MoCE-IR leaves
residual haze, while \method preserves natural tones.
On haze+rain and low+haze+rain, both baselines retain visible streaks
and washed-out colors, whereas \method removes streaks cleanly and
recovers sharper texture.
The error maps mirror these observations: \method's residuals stay
visibly darker, confirming that the quantitative
gains in \Cref{tab:cdd11} translate into perceptually meaningful
improvements.}

\begin{table*}[!t]
\centering
\caption{
  \td{Results on CDD-11 (PSNR / SSIM).
  All methods are trained from scratch on CDD-11.
  L: low-light, H: haze, R: rain, S: snow.
  \emph{Avg} is the macro-average over all $11$ conditions.
  Row colors mark the overall \colorbox{lyellow}{best},
  \colorbox{lblue}{second-best}, and \colorbox{lgray}{third-best}
  methods (ranked by Avg PSNR).}
}
\label{tab:cdd11}
\setlength{\tabcolsep}{4pt}
\resizebox{\linewidth}{!}{%
\begin{tabular}{r *{12}{c}}
\toprule
& \multicolumn{4}{c}{\bf Single degradation}
& \multicolumn{5}{c}{\bf Two degradations}
& \multicolumn{2}{c}{\bf Three degradations}
& \\
\cmidrule(lr){2-5} \cmidrule(lr){6-10} \cmidrule(lr){11-12}
\textbf{Methods}
& {L} & {H} & {R} & {S}
& {L+H} & {L+R} & {L+S} & {H+R} & {H+S}
& {L+H+R} & {L+H+S}
& \textbf{Avg} \\
\midrule
AirNet~\citep{li2022airnet}
  & 24.83 / .778 & 24.21 / .951 & 26.55 / .891 & 26.79 / .919
  & 23.23 / .779 & 22.82 / .710 & 23.29 / .723 & 22.21 / .868 & 23.29 / .901
  & 21.80 / .708 & 22.24 / .725 & 23.75 / .814 \\
PromptIR~\citep{potlapalli2023promptir}
  & 26.32 / .805 & 26.10 / .969 & 31.56 / .946 & 31.53 / .960
  & 24.49 / .789 & 25.05 / .771 & 24.51 / .761 & 24.54 / .924 & 23.70 / .925
  & 23.74 / .752 & 23.33 / .747 & 25.90 / .850 \\
WGWSNet~\citep{zhu2023learning}
  & 24.39 / .774 & 27.90 / .982 & 33.15 / .964 & 34.43 / .973
  & 24.27 / .800 & 25.06 / .772 & 24.60 / .765 & 27.23 / .955 & 27.65 / .960
  & 23.90 / .772 & 23.97 / .771 & 26.96 / .863 \\
WeatherDiff~\citep{ozdenizci2023restoring}
  & 23.58 / .763 & 21.99 / .904 & 24.85 / .885 & 24.80 / .888
  & 21.83 / .756 & 22.69 / .730 & 22.12 / .707 & 21.25 / .868 & 21.99 / .868
  & 21.23 / .716 & 21.04 / .698 & 22.49 / .799 \\
\rowcolor{lgray} OneRestore~\citep{guo2024onerestore}
  & 26.48 / \underline{.826} & 32.52 / \underline{.990} & 33.40 / .964 & 34.31 / .973
  & 25.79 / \underline{.822} & 25.58 / .799 & 25.19 / .789 & \underline{29.99} / .957 & \underline{30.21} / .964
  & 24.78 / .788 & 24.90 / \underline{.791} & 28.47 / .878 \\
\rowcolor{lblue} MoCE-IR~\citep{zamfir2025complexity}
  & \textbf{27.26} / .824 & \underline{32.66} / \underline{.990} & \underline{34.31} / \underline{.970} & \underline{35.91} / \underline{.980}
  & \underline{26.24} / .817 & \underline{26.25} / \underline{.800} & \underline{26.04} / \underline{.793} & 29.93 / \underline{.964} & 30.19 / \underline{.970}
  & \underline{25.41} / \underline{.789} & \underline{25.39} / .790 & \underline{29.05} / \underline{.881} \\
\midrule
\rowcolor{lyellow} \textbf{\method (ours)}
  & \underline{27.05} / \textbf{.836} & \textbf{34.52} / \textbf{.994} & \textbf{35.46} / \textbf{.976} & \textbf{37.54} / \textbf{.983}
  & \textbf{26.40} / \textbf{.836} & \textbf{26.40} / \textbf{.819} & \textbf{26.33} / \textbf{.814} & \textbf{31.61} / \textbf{.974} & \textbf{32.28} / \textbf{.978}
  & \textbf{25.93} / \textbf{.813} & \textbf{25.96} / \textbf{.814} & \textbf{29.95} / \textbf{.894} \\
\bottomrule
\end{tabular}}
\end{table*}

\begin{table*}[!t]
\centering
\caption{
 \td{Results on the conventional \textbf{3-task (\textcolor{purple}{3D})} all-in-one benchmark (PSNR / SSIM).
  Tasks: dehazing on SOTS, deraining on Rain100L, and Gaussian
  denoising on BSD68 at $\sigma\in\{15,25,50\}$.
  Row colors mark the overall \colorbox{lyellow}{best},
  \colorbox{lblue}{second-best}, and \colorbox{lgray}{third-best}
  methods.}
}
\vspace{-1mm}
\label{tab:allinone-3d}
\resizebox{\linewidth}{!}{
\begin{tabular}{r c*{12}{c}}
\toprule
& \multicolumn{2}{c}{\bf Dehazing} & \multicolumn{2}{c}{\bf Deraining} & \multicolumn{6}{c}{\bf Denoising (BSD68~\citep{martin2001database})} & & \\
\textbf{Methods} & \multicolumn{2}{c}{SOTS~\citep{li2018benchmarking}} & \multicolumn{2}{c}{Rain100L~\citep{yang2017deep}} & \multicolumn{2}{c}{$\sigma = 15$} & \multicolumn{2}{c}{$\sigma = 25$} & \multicolumn{2}{c}{$\sigma = 50$} & \multicolumn{2}{c}{\bf Average} \\
\cmidrule(lr){2-3} \cmidrule(lr){4-5} \cmidrule(lr){6-7} \cmidrule(lr){8-9} \cmidrule(lr){10-11} \cmidrule(lr){12-13}
 & {PSNR$\uparrow$} & {SSIM$\uparrow$} & {PSNR$\uparrow$} & {SSIM$\uparrow$} & {PSNR$\uparrow$} & {SSIM$\uparrow$} & {PSNR$\uparrow$} & {SSIM$\uparrow$} & {PSNR$\uparrow$} & {SSIM$\uparrow$} & {PSNR$\uparrow$} & {SSIM$\uparrow$} \\
\midrule
MPRNet~\citep{zamir2021multi}
    & 25.28 & 0.955 & 33.57 & 0.954 & 33.54 & 0.927 & 30.89 & 0.880 & 27.56 & 0.779 & 30.17 & 0.899 \\
Gridformer~\citep{wang2024gridformer}
    & 30.37 & 0.970 & 37.15 & 0.972 & 33.93 & 0.931 & 31.37 & 0.887 & 28.11 & 0.801 & 32.19 & 0.912 \\
Art-PromptIR~\citep{wu2024harmony}
    & 30.83 & 0.979 & 37.94 & 0.982 & 34.06 & 0.934 & 31.42 & \underline{0.891} & 28.14 & 0.801 & 32.48 & 0.917 \\
DA-CLIP~\citep{luo2023controlling}
    & 29.46 & 0.963 & 36.28 & 0.968 & 30.02 & 0.821 & 24.86 & 0.585 & 22.29 & 0.476 & 28.58 & 0.763 \\
UniProcessor~\citep{duan2024uniprocessor}
    & \textbf{31.66} & 0.979 & 38.17 & \underline{0.982} & 34.08 & \underline{0.935} & 31.42 & \underline{0.891} & 28.17 & \underline{0.803} & 32.70 & \underline{0.918} \\
\rowcolor{lgray} PromptIR~\citep{potlapalli2023promptir}
    & 30.58 & 0.974 & 36.37 & 0.972 & 33.98 & 0.933 & 31.31 & 0.888 & 28.06 & 0.799 & 32.06 & 0.913 \\
\rowcolor{lblue} MoCE-IR~\citep{zamfir2025complexity}
    & \underline{31.34} & \textbf{0.979}
    & \underline{38.57} & \textbf{0.984}
    & \underline{34.11} & 0.932
    & \underline{31.45} & 0.888
    & \underline{28.18} & 0.800
    & \underline{32.73} & 0.917 \\
\midrule
\rowcolor{lyellow} \method-\textcolor{purple}{\bf 3D}
    & 30.96 & \underline{0.977}
    & \textbf{38.82} & \textbf{0.984}
    & \textbf{34.45} & \textbf{0.949}
    & \textbf{32.19} & \textbf{0.926}
    & \textbf{29.08} & \textbf{0.877}
    & \textbf{33.10} & \textbf{0.943} \\
\bottomrule
\end{tabular}
}
\end{table*}

\begin{table*}[!t]
\centering
\caption{
 \td{Results on the conventional \textbf{5-task (\textcolor{purple}{5D})} all-in-one benchmark (PSNR / SSIM).
  Tasks: the three 3D tasks, plus deblurring on GoPro, and low-light enhancement
  on LOL-v1.
  Row colors mark the overall \colorbox{lyellow}{best},
  \colorbox{lblue}{second-best}, and \colorbox{lgray}{third-best}
  methods.}
}
\label{tab:allinone-5d}
\setlength{\tabcolsep}{4pt}
\resizebox{\linewidth}{!}{%
\begin{tabular}{r *{12}{c}}
\toprule
& \multicolumn{2}{c}{\bf Dehazing}
& \multicolumn{2}{c}{\bf Deraining}
& \multicolumn{2}{c}{\bf Denoising}
& \multicolumn{2}{c}{\bf Deblurring}
& \multicolumn{2}{c}{\bf Low-Light}
& \multicolumn{2}{c}{\bf Average} \\
\textbf{Methods}
& \multicolumn{2}{c}{SOTS~\citep{li2018benchmarking}}
& \multicolumn{2}{c}{Rain100L~\citep{yang2017deep}}
& \multicolumn{2}{c}{BSD68~\citep{martin2001database} ($\sigma{=}25$)}
& \multicolumn{2}{c}{GoPro~\citep{nah2017deep}}
& \multicolumn{2}{c}{LOLv1~\citep{wei2018deep}}
& \multicolumn{2}{c}{} \\
\cmidrule(lr){2-3} \cmidrule(lr){4-5} \cmidrule(lr){6-7} \cmidrule(lr){8-9} \cmidrule(lr){10-11} \cmidrule(lr){12-13}
 & {PSNR$\uparrow$} & {SSIM$\uparrow$}
& {PSNR$\uparrow$} & {SSIM$\uparrow$}
& {PSNR$\uparrow$} & {SSIM$\uparrow$}
& {PSNR$\uparrow$} & {SSIM$\uparrow$}
& {PSNR$\uparrow$} & {SSIM$\uparrow$}
& {PSNR$\uparrow$} & {SSIM$\uparrow$} \\
\midrule
NAFNet~\citep{chen2022simple}       & 25.23 & .939 & 35.56 & .967 & 31.02 & .883 & 26.53 & .808 & 20.49 & .809 & 27.76 & .881 \\
DGUNet~\citep{mou2022deep}          & 24.78 & .940 & 36.62 & .971 & 31.10 & .883 & 27.25 & .837 & 21.87 & .823 & 28.32 & .891 \\
Restormer~\citep{zamir2022restormer} & 24.09 & .927 & 34.81 & .962 & 31.49 & .884 & 27.22 & .829 & 20.41 & .806 & 27.60 & .881 \\
MambaIR~\citep{guo2024mambair}       & 25.81 & .944 & 36.55 & .971 & 31.41 & .884 & 28.61 & .875 & 22.49 & .832 & 28.97 & .901 \\
Transweather~\citep{valanarasu2022transweather} & 21.32 & .885 & 29.43 & .905 & 29.00 & .841 & 25.12 & .757 & 21.21 & .792 & 25.22 & .836 \\
IDR~\citep{zhang2023ingredient}     & 25.24 & .943 & 35.63 & .965 & \textbf{31.60} & \underline{.887} & 27.87 & .846 & 21.34 & .826 & 28.34 & .893 \\
Gridformer~\citep{wang2024gridformer} & 26.79 & .951 & 36.61 & .971 & 31.45 & .885 & 29.22 & .884 & 22.59 & .831 & 29.33 & .904 \\
\rowcolor{lgray} InstructIR-5D~\citep{conde2024instructir}
             & 27.10 & .956 & 36.84 & .973 & 31.40 & .873 & 29.40 & .886 & \underline{23.00} & .836 & 29.55 & .908 \\
\rowcolor{lblue} MoCE-IR~\citep{zamfir2025complexity}
             & \textbf{30.48} & \underline{.974}
             & \underline{38.04} & \underline{.982}
             & 31.34 & \underline{.887}
             & \textbf{30.05} & \textbf{.899}
             & \underline{23.00} & \underline{.852}
             & \underline{30.58} & \underline{.919} \\
\midrule
\rowcolor{lyellow} \textbf{\method-\textcolor{purple}{\bf 5D} (ours)}
             & \underline{30.35} & \textbf{.976}
             & \textbf{38.87} & \textbf{.984}
             & \underline{31.59} & \textbf{.917}
             & \underline{29.70} & \underline{.895}
             & \textbf{23.87} & \textbf{.861}
             & \textbf{30.87} & \textbf{.927} \\
\bottomrule
\end{tabular}}
\end{table*}

\subsection{Routing behavior analysis}
\label{sec:routing_analysis}
\label{sec:routing_vis}

\begin{figure}[t]
  \centering
  \begin{subfigure}[t]{0.3\linewidth}
    \centering
    \includegraphics[width=\linewidth]{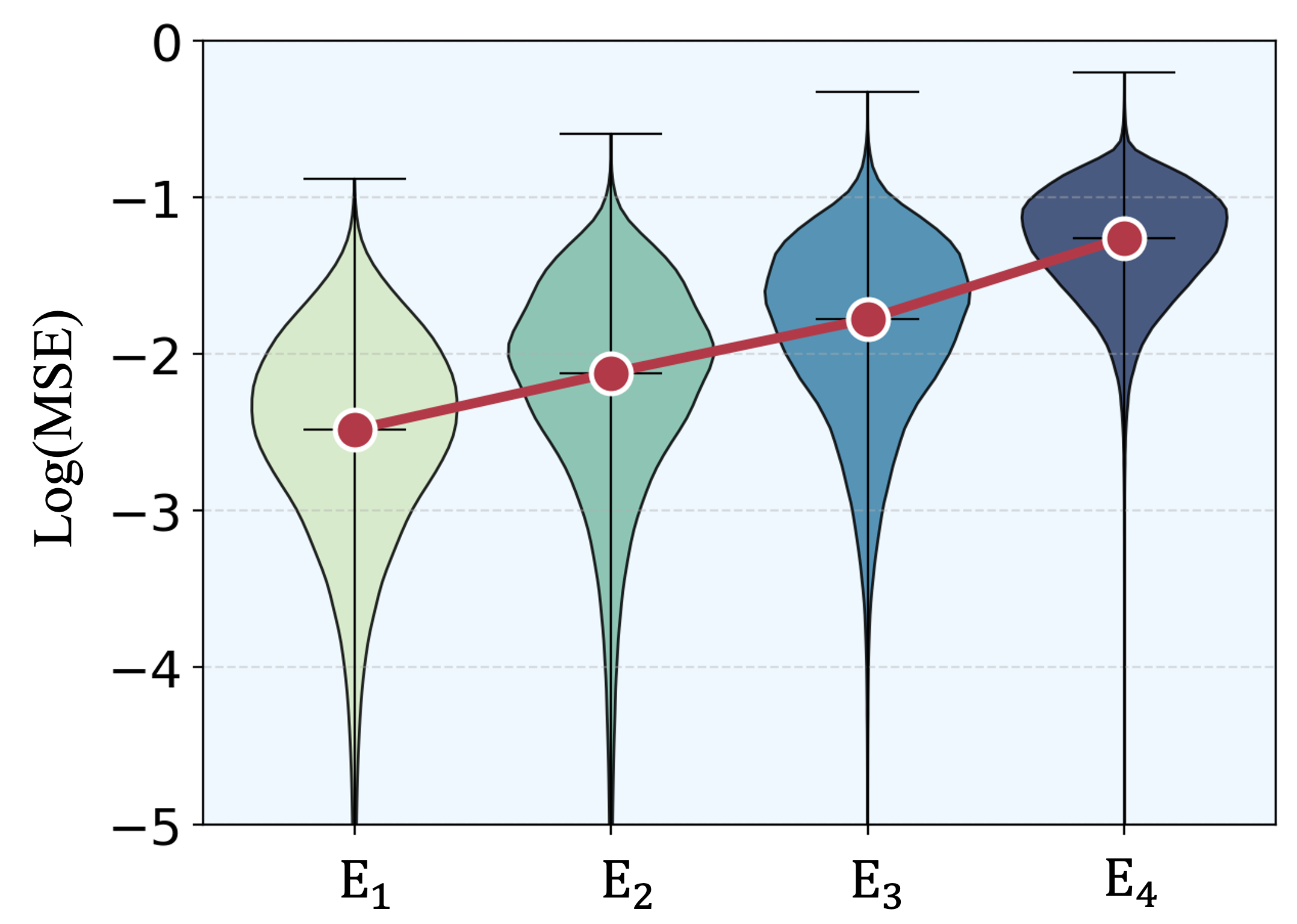}
    \caption{Spatial routing.}
    \label{fig:routing-content}
  \end{subfigure}
  \hfill
  \begin{subfigure}[t]{0.69\linewidth}
    \centering
    \includegraphics[width=\linewidth]{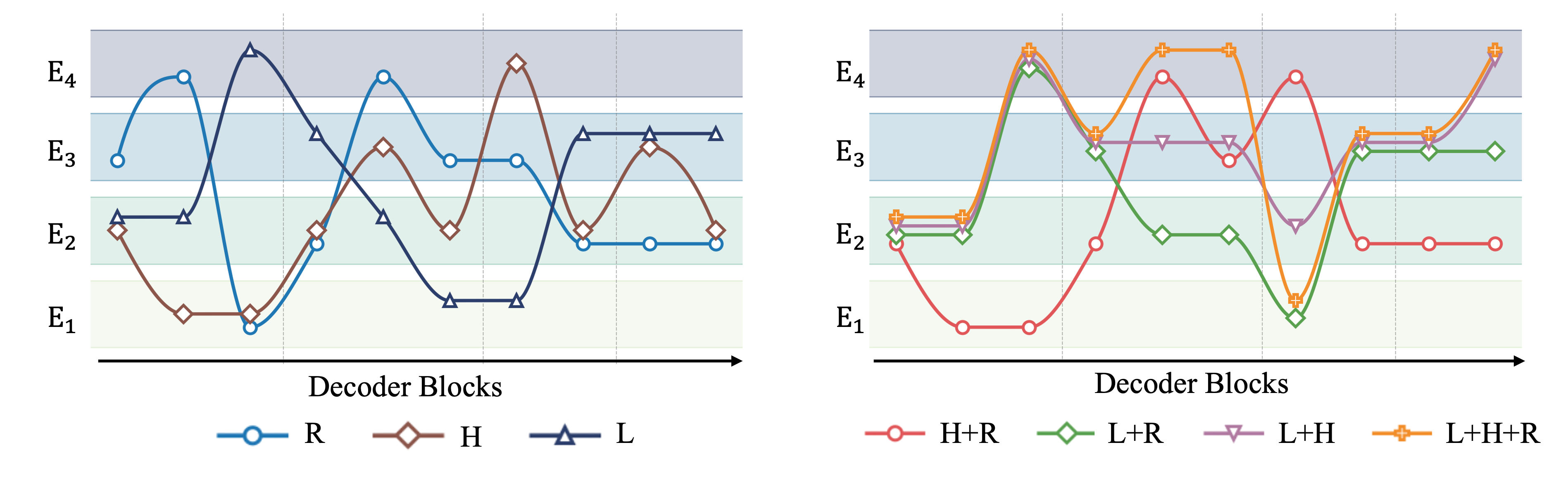}
    \caption{Channel routing: single (left) and composite (right) degradations.}
    \label{fig:routing-task}
  \end{subfigure}
  \caption{%
    \td{Routing behavior on CDD-11.
    \textbf{(a)} Per-patch log MSE grouped by the spatial expert
    selected to process the patch; the mean increases monotonically
    from $E_1$ to $E_4$.
    \textbf{(b)} Most-frequent channel expert at each decoder block,
    one path per degradation type; composite paths largely overlap
    with those of their constituents.}}
  \label{fig:routing_analysis}
\end{figure}

\td{To check whether the two mixers instantiate the intended
\emph{where} / \emph{what} decomposition of \method, we probe whether
the spatial router routes harder patches to higher-capacity experts
and whether the channel router sends different degradations through
distinct decoder paths. \Cref{fig:routing_analysis} reports both
probes on the CDD-11 test set.}

\textbf{Spatial routing tracks patch difficulty.}
\td{We measure per-patch difficulty by the log MSE between a degraded
patch and its target, then group patches by their assigned expert
(\Cref{fig:routing-content}). The mean log MSE increases monotonically
from $E_1$ to $E_4$: easy patches concentrate at the lightweight
$E_1$ while heavily corrupted patches reach the full-capacity $E_4$.
Without any direct supervision on the routing decision, the spatial
router matches patch difficulty to expert capacity, consistent with
the graded-capacity design of \Cref{sec:content}.}

\textbf{Channel routing reflects compositional task identity.}
\td{For each task, we record the most-frequently selected expert at
each channel-mixer block across the test images, yielding one
representative path per task. Single degradations
(\Cref{fig:routing-task}, left) follow well-separated decoder paths,
showing that the multi-hot classification objective distinguishes
elemental tasks. Composite inputs (right) trace paths that largely
overlap with those of their constituents; \eg\ haze+rain blends
routes selected by haze and by rain. The composite routing thus
inherits elemental structure and recombines it at inference,
providing a mechanism for the multi-degradation gains in
\Cref{tab:cdd11}.}

\subsection{Complexity analysis}
\label{sec:complexity}

\begin{figure}[t]
  \centering
  \begin{minipage}[c]{0.36\linewidth}
    \centering
    \includegraphics[width=\linewidth]{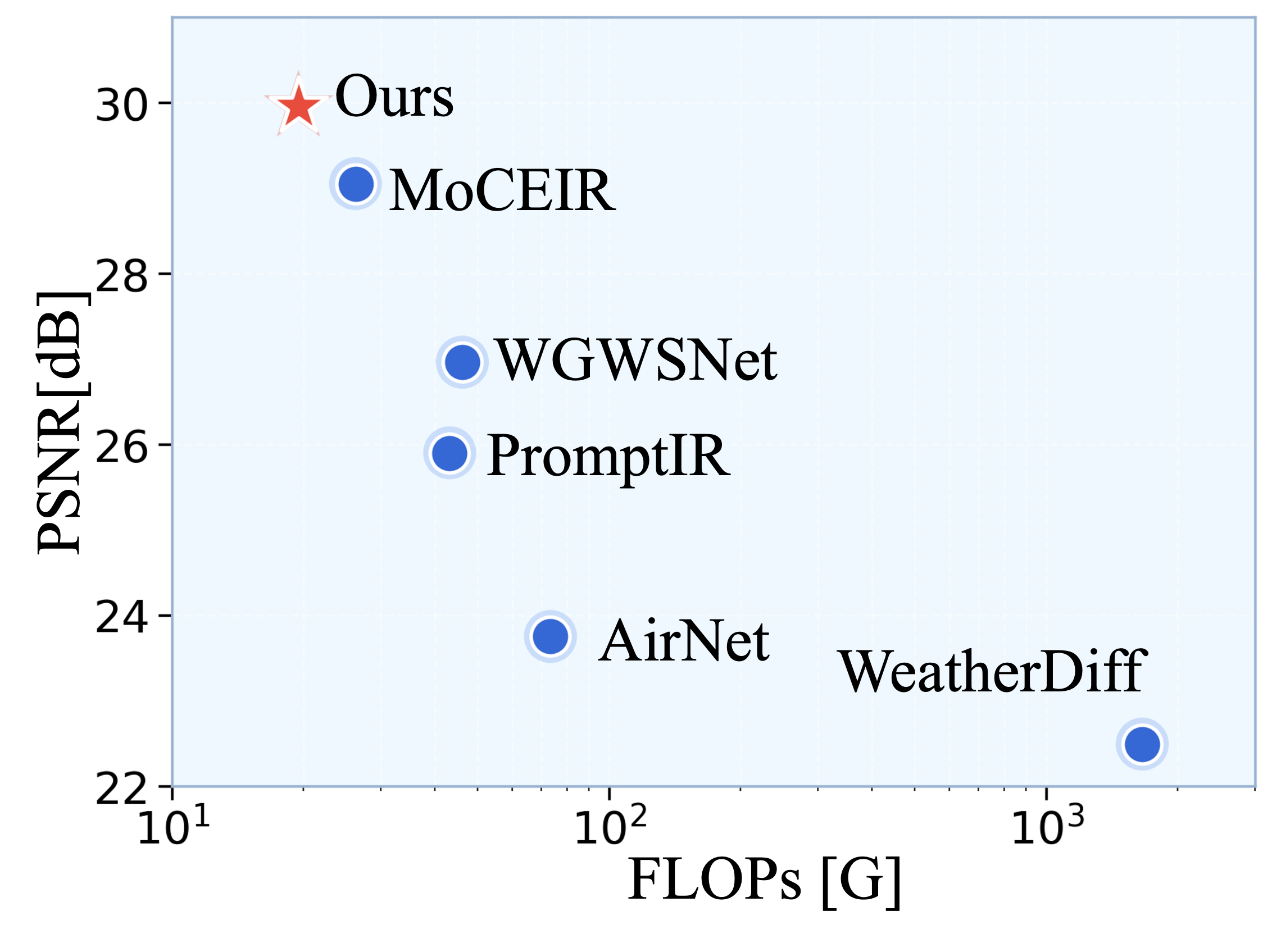}
  \end{minipage}\hfill
  \begin{minipage}[c]{0.60\linewidth}
  \centering
  \small
  \resizebox{\linewidth}{!}{%
  \begin{tabular}{lcccc}
    \toprule
    Method & Params (M) & FLOPs (G) & Mem (MB) & Time (ms) \\
    \midrule
    AirNet~\citep{li2022airnet}                  &  8.93 &   73.42 & 122 &  38.44 \\
    PromptIR~\citep{potlapalli2023promptir}      & 35.59 &   43.18 & 288 &  35.33 \\
    WGWS-Net~\citep{zhu2023learning}             & 19.77 &   46.15 & 113 &  60.71 \\
    WeatherDiff~\citep{ozdenizci2023restoring}   & 85.61 & 1658.45 & 431 & 383.75 \\
    \rowcolor{lblue} MoCE-IR~\citep{zamfir2025complexity}         & 25.35 &   26.29 & 161 &  45.46 \\
    \midrule
    \rowcolor{lyellow} \textbf{\method} (ours) & 32.58 &  \textbf{19.47} & 195 & \textbf{32.11} \\
    \bottomrule
  \end{tabular}
  }
\end{minipage}
  \caption{
    \td{Computational efficiency on CDD-11.
    Left: PSNR vs.\ FLOPs trade-off (log scale); \method occupies the
    upper-left corner. Right: parameters, FLOPs, peak memory, and
    runtime measured on $1{\times}3{\times}128{\times}128$ inputs at
    batch size $1$ on a single A5000 GPU in FP32.}
    }
  \label{fig:efficiency}
\end{figure}

\td{\Cref{fig:efficiency} shows that \method requires only $19.47$~G FLOPs
($26\%$ less than MoCE-IR) and runs the fastest at $32.11$~ms among
all compared methods, while also attaining the highest Avg PSNR on
CDD-11.
The MoE design stores all $K$ experts even though only one is
activated per input, so \method has a higher parameter count
($32.58$~M) and peak memory ($195$~MB) than MoCE-IR; we view this as
a worthwhile trade since compute and latency dominate at inference.}

\subsection{Ablation studies}
\label{sec:ablation}

\td{We ablate three design choices on CDD-11: the multi-hot classification
head, the two routing branches, and the training objective. Each variant
changes only the component under study and shares the backbone, training
schedule, and $K{=}4$ expert configuration with the full model. Scores are
averaged within the single-, two-, and three-degradation subsets.}

\begin{wraptable}{r}{0.58\linewidth}
  \vspace{-12pt}
  \centering
  \caption{{Ablations on CDD-11.}
  Each cell reports PSNR / SSIM averaged over single, two-, and three-degradation conditions; \textit{Avg}
  is the macro-average over all $11$ conditions.}
  \label{tab:ablation}
  \scriptsize
  \setlength{\tabcolsep}{3pt}
  \renewcommand{\arraystretch}{1.05}
  \resizebox{\linewidth}{!}{%
  \begin{tabular}{lcccc}
    \toprule
    Variant & Single & Two & Three & Avg \\
    \midrule
    \method (full)
      & \textbf{33.64 / .947} & \textbf{28.60 / .884}
      & \textbf{25.95 / .814} & \cellcolor{lblue}\textbf{29.95 / .894} \\
    \midrule
    \multicolumn{5}{l}{\textit{Multi-hot classification supervision}} \\
    \;\;Single-label (11-class CE)
      & 32.67 / .945 & 27.83 / .878 & 25.40 / .805 & \cellcolor{lblue}29.15 / .889 \\
    \midrule
    \multicolumn{5}{l}{\textit{Spatial/channel mixer design}} \\
    \;\;w/o spatial routing
      & 33.59 / .948 & 28.58 / .884 & 25.86 / .812 & \cellcolor{lblue}29.90 / .894 \\
    \;\;w/o channel routing
      & 33.24 / .947 & 28.15 / .881 & 25.72 / .809 & \cellcolor{lblue}29.56 / .892 \\
    \;\;w/o both routing
      & 32.95 / .946 & 28.07 / .882 & 25.79 / .812 & \cellcolor{lblue}29.43 / .893 \\
    \midrule
    \multicolumn{5}{l}{\textit{Training objective}} \\
    \;\;w/o $\calL_{\mathrm{task}}$
      & 33.14 / .944 & 28.19 / .878 & 25.48 / .803 & \cellcolor{lblue}29.49 / .889 \\
    \;\;w/o $\calL_{\mathrm{lb}}$
      & 32.88 / .947 & 28.14 / .883 & 25.69 / .812 & \cellcolor{lblue}29.42 / .893 \\
    \;\;w/o $\calL_{\mathrm{cdrl}}$
      & 33.15 / .946 & 28.28 / .881 & 25.73 / .809 & \cellcolor{lblue}29.59 / .892 \\
    \;\;w/o $\calL_{\mathrm{ms}}$
      & 33.39 / .947 & 28.55 / .884 & 25.77 / .813 & \cellcolor{lblue}29.80 / .894 \\
    \bottomrule
  \end{tabular}}
  \vspace{-10pt}
\end{wraptable}

\textbf{Multi-hot classification supervision.}
\td{To check whether composite task identity benefits from a compositional
representation, we replace the multi-hot BCE head with an $11$-class
softmax that treats each degradation combination as a separate label.
The Avg PSNR drops from $29.95$ to $29.15$~dB ($-0.80$~dB), the largest
single drop in the ablation, with a wider gap on three-way mixtures
($-0.55$~dB). Multi-hot supervision lets elemental components be shared
across mixtures, whereas the single-label head forces each combination to
be learned in isolation.}

\textbf{Spatial / channel mixer design.}
\td{To isolate the two adaptive decisions, we remove spatial  and
channel routing separately. For fair comparison, each removed branch
(expert widths $[12,24,36,48]$) is replaced by a single FLOPs-equivalent
expert with $d{=}32$. Removing channel routing causes the larger drop
($-0.39$~dB on Avg), confirming the dominant role of explicit task-aware
routing; removing spatial routing accounts for a $-0.05$~dB drop.
Removing both drops Avg by $-0.52$~dB, slightly above the sum of
the individual effects, so the two routers contribute complementary
signals and jointly support the \emph{where} / \emph{what} design.}

\textbf{Training objective.}
\td{To assess each auxiliary loss, we remove one term at a time and keep
the rest of the objective unchanged. Removing $\calL_{\mathrm{lb}}$
($-0.53$~dB) or $\calL_{\mathrm{task}}$ ($-0.46$~dB) causes the largest
drops, confirming that both balanced expert usage and explicit multi-hot
supervision are central to the routing mechanism. The contrastive term
$\calL_{\mathrm{cdrl}}$ ($-0.36$~dB) contributes by sharpening
task-specific reconstructions, and the multi-scale SSIM term
$\calL_{\mathrm{ms}}$ ($-0.15$~dB) plays a smaller but still positive role.}

\section{Conclusion}
\label{sec:conclusion}

\td{We introduced \method, a composite image restoration framework that jointly
decides \emph{where} to spend computation and \emph{what} degradation types
to specialize for. By coupling patch-level spatial routing with multi-hot
task-conditioned channel routing, \method achieves state-of-the-art
performance on CDD-11 and on the conventional 3-task and 5-task all-in-one
benchmarks, while remaining the most efficient among compared methods in
FLOPs and runtime. Promising directions for future work include broader
evaluation on real-world images with perceptual studies, improving expert
efficiency for real-time use, and extending \method to video restoration
with temporally stable routing.}

\bibliographystyle{plainnat}
\bibliography{refs}

\newpage
\appendix

\section{Detailed Architecture}
\label{sec:app_arch}

This appendix provides implementation details that complement the main method in \Cref{sec:method} and the overview in \Cref{fig:overview}.
We fix base channel dimension $d=48$ and $K=4$ experts throughout; all configurations scale proportionally across the four encoder-decoder levels.

\subsection{Encoder-Decoder Configuration}

The U-Net has four levels with channel dimensions $[48,\,96,\,192,\,384]$, attention head counts $[1,\,2,\,4,\,8]$, and block counts $[2,\,3,\,3,\,4]$.
Each block alternates a channel mixer and a spatial mixer.
Downsampling is performed by a $3\!\times\!3$ convolution followed by PixelUnshuffle($2$); upsampling uses a $3\!\times\!3$ convolution followed by PixelShuffle($2$).
Skip connections from encoder to decoder are concatenated and fused with a $1\!\times\!1$ convolution.

\subsection{Attention Mechanisms}
\label{sec:app_attn}

Both the spatial mixer and channel mixer are built on attention, but they operate along different dimensions.
This subsection expands the attention operators used by the two mixers introduced in \Cref{sec:content,sec:task}.

\paragraph{Window-based spatial attention.}
The spatial attention module operates within non-overlapping $q \times q$ windows. Given an input $\mX \in \R^{N \times C}$ where $N = q^2$ is the number of tokens per window, queries, keys, and values are computed by a single linear projection and split across $h$ attention heads with head dimension $d_h = C / h$:
\e
\mQ, \mK, \mV = \mrm{split}(\mW_{\mrm{qkv}}\,\mX), \qquad \mQ, \mK, \mV \in \R^{h \times N \times d_h}.
\ee
The attention matrix incorporates a \emph{continuous relative position bias} (CPB)~\cite{liu2022swin}. Rather than storing a fixed bias table, a small MLP ($2 \to 512 \to h$) maps normalized relative coordinates to per-head biases, allowing generalisation to unseen window sizes:
\e
\mrm{Attn} = \mrm{Softmax}\!\left(\frac{\mQ \mK^\T}{\sqrt{d_h}} + \mB_{\mrm{CPB}}\right)\mV,
\ee
where $\mB_{\mrm{CPB}} \in \R^{h \times N \times N}$ is the position bias output of the CPB-MLP indexed by pairwise relative positions. The output is projected back to $C$ dimensions via a linear layer.

Each \texttt{SpatialExpert} wraps this attention module at a configurable hidden dimension $d_k$. When $d_k \neq C$, linear projections $\mW_{\mrm{in}} \in \R^{C \times d_k}$ and $\mW_{\mrm{out}} \in \R^{d_k \times C}$ map features into and out of the expert's subspace; when $d_k = C$ these projections are omitted.

\paragraph{Transposed channel attention.}
The channel attention module follows the multi-DConv head transposed attention (MDTA) design~\cite{zamir2022restormer}. Unlike spatial attention which computes affinities between spatial tokens, MDTA transposes the attention to operate along the \emph{channel} dimension. Given an input $\mX \in \R^{C \times HW}$, queries, keys, and values are generated by a $1\!\times\!1$ convolution followed by a $3\!\times\!3$ depth-wise convolution:
\e
\mQ, \mK, \mV = \mrm{split}\!\left(\mrm{DWConv}_{3\times3}(\mrm{Conv}_{1\times1}(\mX))\right), \qquad \mQ, \mK, \mV \in \R^{h \times \frac{C}{h} \times HW}.
\ee
Queries and keys are $\ell_2$-normalized along the spatial dimension, and a learnable temperature $\valpha \in \R^{h}$ scales the attention logits:
\e
\mrm{Attn} = \mrm{Softmax}\!\left(\hat{\mQ}\,\hat{\mK}^\T \cdot \valpha\right)\mV, \qquad \hat{\mQ} = \frac{\mQ}{\|\mQ\|_2}, \;\; \hat{\mK} = \frac{\mK}{\|\mK\|_2},
\ee
where the matrix product $\hat{\mQ}\,\hat{\mK}^\T \in \R^{h \times \frac{C}{h} \times \frac{C}{h}}$ captures cross-channel correlations. The output is reshaped to spatial layout and projected by a $1\!\times\!1$ convolution. Because the attention dimension is $C/h$ rather than $HW$, this module has complexity $\calO(C^2 HW / h)$ instead of the $\calO(CH^2W^2)$ of standard spatial attention, making it efficient for high-resolution features.

Each \texttt{ChannelExpert} wraps this MDTA module at a configurable hidden dimension $d_k$ with the same projection strategy as the spatial experts.

\paragraph{Feed-forward network.}
Both mixer blocks share the same gated depth-wise feed-forward network (GDFN)~\cite{zamir2022restormer}. A $1\!\times\!1$ convolution expands the channel dimension by a factor $r$ into two branches, a $3\!\times\!3$ depth-wise convolution enriches local context, and a gating mechanism modulates the output:
\e
\mrm{GDFN}(\mX) = \mW_{\mrm{out}}\!\left(\mrm{GELU}(\vx_1) \odot \vx_2\right), \qquad [\vx_1, \vx_2] = \mrm{DWConv}_{3\times3}(\mW_{\mrm{in}}\,\mX),
\ee
where $\mW_{\mrm{in}} \in \R^{C \times 2rC}$, $\mW_{\mrm{out}} \in \R^{rC \times C}$, and $\odot$ denotes element-wise multiplication. We use expansion ratio $r = 2$ throughout.

\subsection{Spatial Mixer Details}
\label{sec:app_spatial}

\paragraph{Content feature module.}
This paragraph specifies the content feature module used by the content router in \Cref{sec:content}.
The global content descriptor $\vz_{\mrm{global}}$ is extracted from the first-level encoder feature rather than from the deeper feature being routed.
Let $\mF_1\in\R^{B\times d\times H\times W}$ be the feature after the overlapped patch embedding.
We apply a lightweight convolutional predictor
\e
\mG = \sigma\!\left(\mrm{Conv}_{3\times3}^{8\to 2}
      \bigl(\sigma(\mrm{Conv}_{1\times1}^{d\to 8}(\mF_1))\bigr)\right)
      \in \R^{B\times 2\times H\times W},
\ee
where $\sigma$ is LeakyReLU with slope $0.1$.
At each encoder, bottleneck, decoder, and refinement stage, $\mG$ is bilinearly resized to the current feature resolution and partitioned into the same non-overlapping $q\times q$ windows as the residual stream.
This gives a per-token content cue $\mG_w\in\R^{q^2\times 2}$ for every routed window, preserving low-level structural information that may be attenuated in deeper layers.

\paragraph{Content router.}
This paragraph gives the implementation corresponding to the router input triplet and logits in \Cref{eq:router}.
For a current feature map $\mX\in\R^{B\times C\times H\times W}$, the spatial mixer first applies LayerNorm and reshapes each $q\times q$ window into token form $\mX_w\in\R^{q^2\times C}$.
The router also uses a fixed two-dimensional coordinate grid $\mP_w\in[-1,1]^{q^2\times 2}$ inside each window.
For every window, the implementation forms the same three inputs used in the main text by averaging the corresponding token-level signals:
\e
\begin{aligned}
\vz_{\mrm{local}}
&= \frac{1}{q^2}\sum_{n=1}^{q^2}\mX_w(n)
   \in\R^{C_\ell},\\
\vz_{\mrm{global}}
&= \frac{1}{q^2}\sum_{n=1}^{q^2}\mG_w(n)
   \in\R^{C_g},\\
\vz_{\mrm{pos}}
&= \frac{1}{q^2}\sum_{n=1}^{q^2}\mP_w(n)
   \in\R^2.
\end{aligned}
\ee
where $C_\ell=C$ and $C_g=2$ in our implementation.
A two-layer router then predicts the expert logits using the same notation as \Cref{eq:router}:
\e
\vl
= \mW_2\,\mathrm{GELU}\!\left(
    \mW_1\,\mathrm{LN}
    ([\vz_{\mrm{local}},\vz_{\mrm{global}},\vz_{\mrm{pos}}])
  \right)
\in \R^K,
\ee
with hidden width $\max(8,(C_\ell+C_g+2)/4)$ in our implementation.
During training, hard Gumbel-Softmax produces a one-hot expert assignment while keeping the logits differentiable.
At inference, the router applies softmax followed by hard $\argmax_k l_k$, groups windows by selected expert, processes only the assigned windows with that expert, and scatters the outputs back to the original window order.
The patch load-balancing loss follows the Switch formulation, using the mean hard assignment as load and the mean softmax probability as importance.

\subsection{Channel Mixer Details}
\label{sec:app_channel}

\paragraph{Task feature module.}
This paragraph details the task feature module described in \Cref{sec:task}; the resulting feature $\vg$ is the representation visualized in \Cref{fig:pca}.
The global task feature $\vg$ is extracted once from the bottleneck feature map $\mZ_{\mrm{btl}}\in\R^{B\times C_{\mrm{btl}}\times H'\times W'}$ and is reused by all routed channel mixers in the decoder and refinement stages.
The module has three complementary branches:
\begin{enumerate}[nosep,leftmargin=1.5em]
    \item \textbf{High-frequency branch.} A frozen depth-wise $3\!\times\!3$ Laplacian filter with kernel $[-1,-1,-1;\;-1,8,-1;\;-1,-1,-1]$ extracts edge and texture responses, followed by GELU, spatial mean pooling, and a two-layer MLP $C_{\mrm{btl}}\to 2C_{\mrm{btl}}\to C_{\mrm{btl}}$.
    \item \textbf{Low-frequency branch.} A $3\!\times\!3$ average pool with stride $1$ and padding $1$ smooths the bottleneck feature, followed by a $1\!\times\!1$ convolution, GELU, and spatial mean pooling.
    \item \textbf{Statistics branch.} Channel-wise mean and standard deviation are concatenated into a $2C_{\mrm{btl}}$-dimensional vector and passed through a two-layer MLP $2C_{\mrm{btl}}\to 2C_{\mrm{btl}}\to C_{\mrm{btl}}$.
\end{enumerate}
The three branch outputs are concatenated and fused as
\e
\vg
= \mW_g^{(2)}\,\mathrm{GELU}\!\left(
   \mW_g^{(1)}\,\mathrm{LN}
   ([\vz_{\mrm{high}},\vz_{\mrm{low}},\vz_{\mrm{stat}}])
  \right)
\in \R^{C_{\mrm{btl}}},
\ee
where $\mW_g^{(1)}:3C_{\mrm{btl}}\to 2C_{\mrm{btl}}$ and
$\mW_g^{(2)}:2C_{\mrm{btl}}\to C_{\mrm{btl}}$.

\paragraph{Multi-hot classification head and task router.}
This paragraph expands the multi-hot task supervision and image-level channel routing introduced in \Cref{sec:task}.
A linear classification head maps $\vg$ to $M$ degradation logits:
\e
\hat{\vy} = \mrm{Sigmoid}(\mW_{\mrm{cls}}\vg) \in [0,1]^M,
\ee
and the head is supervised by binary cross-entropy with the multi-hot degradation label.
In our CDD-11 setting, $M=4$ for rain, snow, haze, and low-light bases.
For channel routing, each routed decoder or refinement block concatenates $\vg$ with the global-average-pooled current feature map and feeds the result to the same LayerNorm $\to$ Linear $\to$ GELU $\to$ Linear router used by the spatial mixer.
Training uses hard Gumbel-Softmax and the corresponding image-level load-balancing loss; inference uses hard $\argmax$ and routes each image to a single channel expert.

\subsection{Training Losses}
\label{sec:app_losses}

\paragraph{MS-SSIM loss.}
This paragraph expands the structural term $\calL_{\mrm{ms}}$ in the total objective \Cref{eq:total_loss}.
Given the restored image $\hat{I}$ and clean target $J$, we use the differentiable multi-scale SSIM loss
\e
\calL_{\mrm{ms}}(\hat{I},J)
= 1 -
\prod_{s=1}^{S-1} \mathrm{CS}_s(\hat{I},J)^{\alpha_s}
\cdot
\mathrm{SSIM}_S(\hat{I},J)^{\alpha_S},
\ee
where $S=5$ and
$\valpha=[0.0448,\,0.2856,\,0.3001,\,0.2363,\,0.1333]$.
At each scale, SSIM is computed with an $11\times 11$ Gaussian window:
\e
\mathrm{SSIM}(x,y)
=
\frac{(2\mu_x\mu_y+c_1)(2\sigma_{xy}+c_2)}
     {(\mu_x^2+\mu_y^2+c_1)(\sigma_x^2+\sigma_y^2+c_2)},
\qquad
\mathrm{CS}(x,y)
=
\frac{2\sigma_{xy}+c_2}{\sigma_x^2+\sigma_y^2+c_2}.
\ee
Images are downsampled by $2\times$ average pooling between scales, matching the implementation.

\paragraph{Composite-degradation contrastive loss.}
This paragraph gives the implementation of $\calL_{\mrm{cdrl}}$ used in \Cref{sec:loss}.
Let $I$ be the degraded input, $J$ the clean positive target, $\hat{I}$ the restored output, and $\{I_o\}_{o=1}^{O}$ the other degraded negatives.
For a perceptual feature extractor $V$, we use the VGG-16 feature layers $\mathcal{K}=\{3,8,15\}$ when available; otherwise the same expression falls back to pixel space.
The contrastive loss is
\e
\calL_{\mrm{cdrl}}
=
\frac{1}{|\mathcal{K}|}
\sum_{k\in\mathcal{K}}
\frac{
  \|V_k(\hat{I}) - V_k(J)\|_1
}{
  \frac{1}{O+1}
  \left(
    \|V_k(\hat{I}) - V_k(I)\|_1
    + \sum_{o=1}^{O}
      \|V_k(\hat{I}) - V_k(I_o)\|_1
  \right)
  + \epsilon
},
\ee
with $\epsilon=10^{-7}$.
This pulls the restoration toward the clean target while keeping it separated from the input degradation and the other degraded negatives.

\paragraph{Load-balancing loss.}
This paragraph expands the load-balancing term $\calL_{\mrm{lb}}$ used by both routers in \Cref{sec:content,sec:task} and summarized in \Cref{sec:loss}.
For each routed layer $r$, let $p_{i,k}^{(r)}$ be the softmax probability that routed item $i$ selects expert $k$, and let $a_{i,k}^{(r)}$ be the corresponding hard one-hot assignment from Gumbel-Softmax during training.
The per-layer Switch-style loss is
\e
\calL_{\mrm{lb}}^{(r)}
=
K\sum_{k=1}^{K}
\underbrace{
  \frac{1}{N_r}\sum_{i=1}^{N_r} a_{i,k}^{(r)}
}_{\mrm{load}_k^{(r)}}
\underbrace{
  \frac{1}{N_r}\sum_{i=1}^{N_r} p_{i,k}^{(r)}
}_{\mrm{importance}_k^{(r)}} ,
\ee
where $N_r$ is the number of routed windows for a spatial mixer or the batch size for a channel mixer.
We average this loss over routed spatial stages and routed channel stages, then average the two available groups to obtain $\calL_{\mrm{lb}}$ used in \Cref{eq:total_loss}.

\end{document}